\documentclass[prl,aps,twocolumn]{revtex4}
\usepackage{amsmath,bm,epsfig}
 \usepackage[latin1]{inputenc}

\newcommand{\K} {_{\rm K}}\newcommand{\N} {_{\rm N}}
 
\def\Fbox#1{\vskip1ex\hbox to 8.5cm{\hfil\fboxsep0.3cm\fbox{%
  \parbox{8.0cm}{#1}}\hfil}\vskip1ex\noindent}  

\let \ve \varepsilon

\def\r{\bm r}

\def\yp {y^+} \def\Kp {K^+} \def\Sp {S^+}
 \def\Wp {W^+}

\let\*\cdot 
\let\=\equiv 
\def\<{\left\langle} \def\>{\right\rangle} \def\({\left(} \def\){\right)}

\let\~\widetilde
\let\^\widehat \def\ort#1{\^{\bf{#1}}}
 \def\x{\ort x} \def\y{\ort y}
\def\z{\ort z}  \def\1{\bm1} 
 
\newcommand{\B}[1]{{\bm{#1}}}
\newcommand{\C}[1]{{\mathcal{#1}}}    
\def\BE{\begin{equation}}\def\EE{\end{equation}}
\def\BEA{\begin{eqnarray}}\def\EEA{\end{eqnarray}}
\def\BSE{\begin{subequations}}\def\ESE{\end{subequations}}
\def \vK {von-K\'arm\'an~}
\def\Re{${{\C R}\mkern-3.1mu e}_\lambda$} 
\def\RE{{{\C R}\mkern-3.1mu e}_\lambda} 

\begin{document}
\title{Estimating von-K\'arm\'an's constant from Homogeneous Turbulence}
\author{T.S. Lo, Victor S. L'vov, Anna Pomyalov  and Itamar Procaccia}
\affiliation{Department of Chemical Physics, The Weizmann Institute
of Science, Rehovot 76100, Israel}

\begin{abstract}
A celebrated universal aspect of wall-bounded turbulent flows is the von 
K\'arm\'an log-law-of-the-wall, describing how the mean velocity in the 
streamwise direction depends on the distance from the wall. Although the 
log-law is known for more than 75 years, the von K\'arm\'an constant 
governing the slope of the log-law was not determined theoretically. In 
this Letter we show that the von-K\'arm\'an constant can be estimated 
from homogeneous turbulent data, i.e. without information from 
wall-bounded flows.
\end{abstract}
\maketitle
 
 \noindent
  \textbf{Introduction.} The theoretical understanding of wall-bounded 
turbulent flows lags behind homogeneous turbulence, in which a number of 
universal constants and exponents can be estimated rather accurately on 
the basis of approximate arguments of considerable success \cite{Fri} . One glaring 
such example of a lack of theoretical power concerns the apparently 
universal log law-of-the-wall which was discovered by von K\'arm\'an in 
1930~\cite{79MY,Pope}. The law pertains to the mean   velocity profile 
$\B V (\r)=  \x \, V (\y \, y )$ in wall bounded  Newtonian turbulence  
($\x$, $\y$ and $\z$  are unit vectors in  the stream-wise, wall-normal and 
span-wise directions respectively). In wall units the law is written as
\BE\label{Karman}
V^+(y^+) = \frac 1 {\kappa\K} \ln y^+ + B\,, \ \mbox{for}\
 30 \le y^+ \ll Re_\lambda  \ . 
\EE
Here the Reynolds number \Re, the normalized distance from 
the wall $y^+$,  and the normalized mean velocity $V^+(y^+)$ (which is 
in the $\x$ direction with a dependence on $y$ only) are defined by
\begin{equation}
 \RE \equiv \frac{L\sqrt{\mathstrut p' L}}{\nu_0}\,, \quad y^+
\equiv \frac{y \RE } L \,, \quad
 V^+ \equiv
\frac V {\sqrt{\mathstrut p'L}}\ .  \label{red}
\end{equation}
Here   $p'$ be the fixed pressure gradients $p'\equiv -\partial 
p/\partial x$,    $\nu_0$   the kinematic viscosity.    The  law 
(\ref{Karman}) is universal, independent of $\RE$, the nature of the 
Newtonian fluid and of the flow geometry over a smooth surface, providing 
that    \Re~ is large enough.

It is one of the shortcomings of the theory of wall-bounded turbulence 
that the von K\'arm\'an constant $\kappa\K= 0.44 \pm  0.03$ and the 
intercept $B\approx 6.13$ are only known from experiments and 
simulations \cite{79MY,97ZS}. In this Letter we propose that $\kappa\K $ 
can be estimated using universal constants that appear in homogenous 
turbulence. As such, we can draw on the relative power of 
homogeneous turbulence theory to improve our understanding of the 
characteristics of wall-bounded flows. We will not draw on any experimental 
information about wall-bounded flows.

In constructing our argument we rely heavily on known facts, including 
recent results, concerning homogeneous isotropic turbulence and 
homogeneous anisotropic turbulence with a constant shear: 
$\partial V (y)/\partial y=  S$. Due to Galilean invariance the 
statistics of the turbulent velocity field ${\B u}(\r,t)$ (from which the
mean $\<\B u(\r,t)\>$ is subtracted) are independent of position.  In 
other words, in such homogeneous and anisotropic ensemble
all statistical object computed at any point, like  the density of the kinetic energy $K$ and  
the Reynolds stress $W$, 
\BE \label{KE}
K\= \<|\B u (\r,t)|^2 \>/2\,, \quad W\=- \<u_x (\r,t)u_y (\r,t)\> \,,
\EE
are space independent. On the other hand, two-point correlation functions, like
the second order longitudinal and transverse structure functions,
\BEA\label{LS2}
S_2(\B r) &\equiv& \big\langle \big( u_{\parallel}(\B r'+\B r,t) -\B 
u_{\parallel}(\B r',t) \big)  ^2\big\rangle\,, \\ \label{TS2}
\~S_2(\B r) &\equiv& \big\langle ( u_{\perp}(\B r'+\B r,t) - 
u_{\perp}(\B r',t) \big)^2\big\rangle\,,
 \EEA
depend only on the vector separation $\r$. In Eq.~(\ref{LS2}) and (\ref{TS2}) $u_{\parallel}$ and 
$u_{\perp}$ are components of $\B u$, parallel and orthogonal to $\r$.  
The physical reason for the homogeneity of the turbulent statistics is that 
the energy flux $\ve$ generated by the pressure head 
\BE\label{input}
\ve= S\, W
\EE
is independent of the mean velocity itself (again, due to Galilean
invariance) and is determined  only by the 
space independent shear $S$ and Reynolds stress $W$. 

\noindent \textbf{I. Similarity of  wall-bounded and constant-shear 
turbulence.} We base our argument on the realization that wall-bounded 
turbulence in the log-law region and 
constant-shear turbulence are very similar.   To be more precise,
constant-shear homogeneous turbulence serves as a  very good approximation to
wall-bounded turbulence;  various characteristics of 
turbulent statistics appear to coincide in the two flows within the available
accuracy of physical experiments and numerical simulations. The basic 
reason for this similarity is precisely that the rate of the energy 
production ~(\ref{input}) depends on the shear itself and not on its space 
derivatives,  which obviously differ in these two flows. The cenral point of this Letter is that the similarity between  wall-bounded and 
constant-shear turbulence allows one to estimate the \vK constant  for 
wall-bounded turbulence using information from homogeneous
constant-shear turbulence.  

The first result that we quote is long 
standing, stating a universal relation between $W$ and $K$ in a 
constant-shear flow,
\begin{equation} \label{defcn}
  W \big / K \=  \Wp \big / \Kp =c^2\N \ , \quad c\N \approx 0.53 \ . 
\end{equation}
The the same value of $c\N$, within the available accuracy (of 
about 5\%), is measured also in the outer layer of channel 
flow~\cite{DNS}. This serves as additional support for the similarity between 
these two types of turbulence. 

Next we use the exact balance equation of mechanical momentum in a 
channel geometry,
\begin{equation}\label{bal1}
\nu_0 S(y) +W(y) = p'(L-y) \ .
\end{equation}
In wall units~(\ref{red}) this equation reads
\begin{equation}\label{bal2}
S^+(\yp) +W^+ (\yp) =1 - \yp/\RE\ .
\end{equation}
For $\RE\gg 1$ the mean shear 
$\Sp$ in the log-law region~(\ref{Karman})  is governed by 
\BE\label{shear+}
\Sp (\yp)=   1\big / ( \kappa\K \, \yp )\,,
\EE
For large values of $\yp$ such that $\yp \ll \RE$  Eq.~(\ref{bal2}) reduces  to
\BE \label{bal3}
\Wp=1\,,
\EE
meaning that in the log-law region  the total momentum flux toward the wall 
is independent of the distance to the wall and is entirely accounted for by
turbulent fluctuations (i.e. the Reynolds stress).
Equations~(\ref{defcn}) and (\ref{bal3}) show that in this region the 
kinetic energy is independent of the distance to the wall and has a
universal value
\BE \label{res1}
\Kp= c\N^{-2}\ .
\EE
This constant value is shared by wall-bounded  and   constant-shear  turbulence.

\noindent\textbf{II. Anisotropy in wall-bounded and  constant-shear 
turbulent flows.}
The knowledge of the total kinetic energy density~(\ref{res1}) is not sufficient
for our purposes, we need to know the distribution of $K $ 
between the different components of turbulent velocity, i.e. the  values of 
\BE \label{defKii}
K_{x}\=  \< u_x^2\>/2 \,, \quad K_{y}\=  \< u_y^2\>/2\,, \quad K_{z}\=  
\< u_z^2\>/2 \ .
\EE

\begin{figure}
\includegraphics[width=0.48 \textwidth]{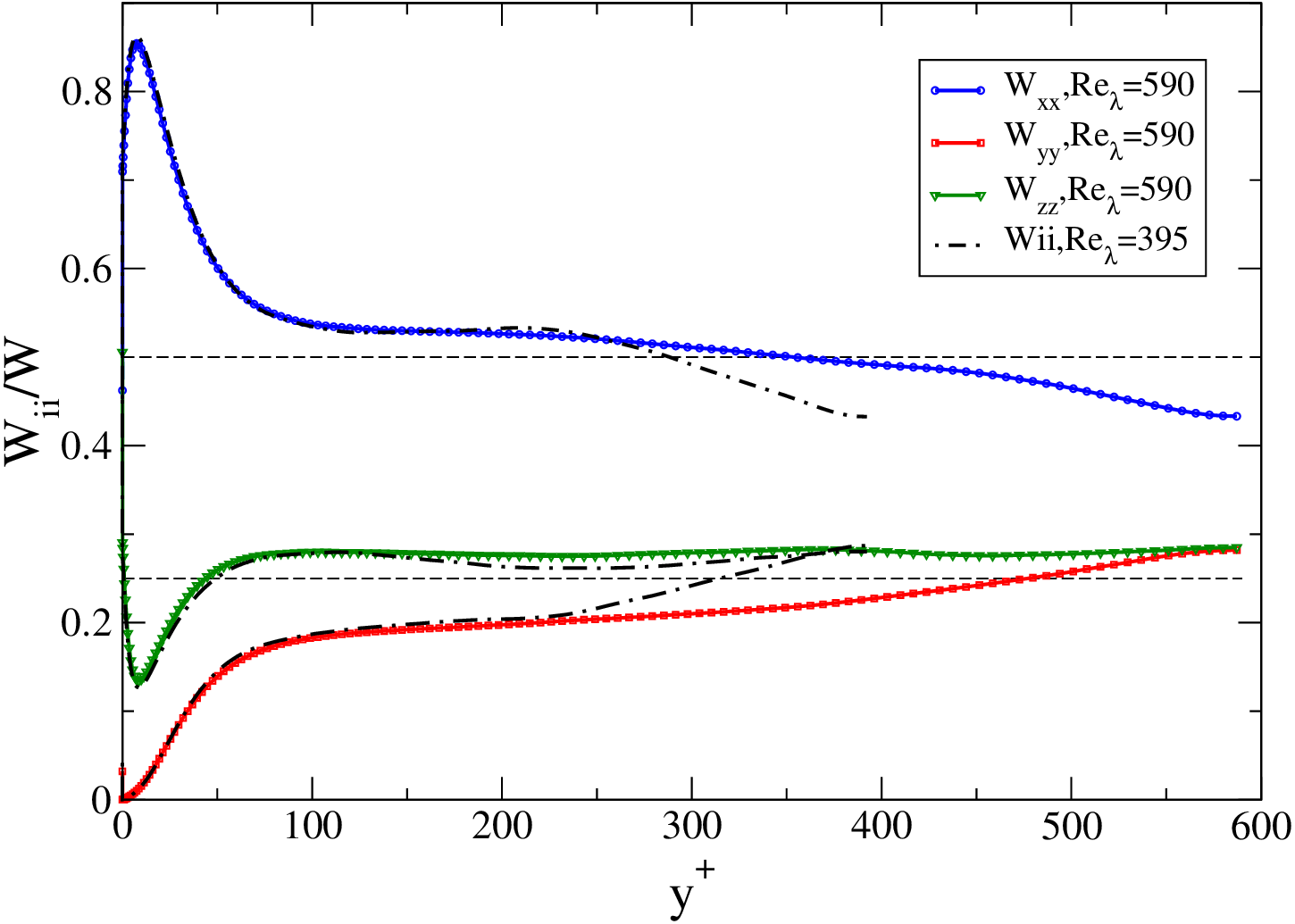}
\caption{\label{f:DNS} DNS profiles of the relative kinetic energies in 
the streamvise direction,  $K_x/K$, lines denoted as ``$xx$",  
wall-normal, $K_y/K$,  and spanwise, $K_z/K$, directions denoted as 
``$yy$" and ``$zz$".  Solid lines:  $\RE =590$, dot-dashed lines: $\RE 
=395$.}
\end{figure}

\begin{figure}
 \centering\includegraphics[width=0.48 \textwidth]{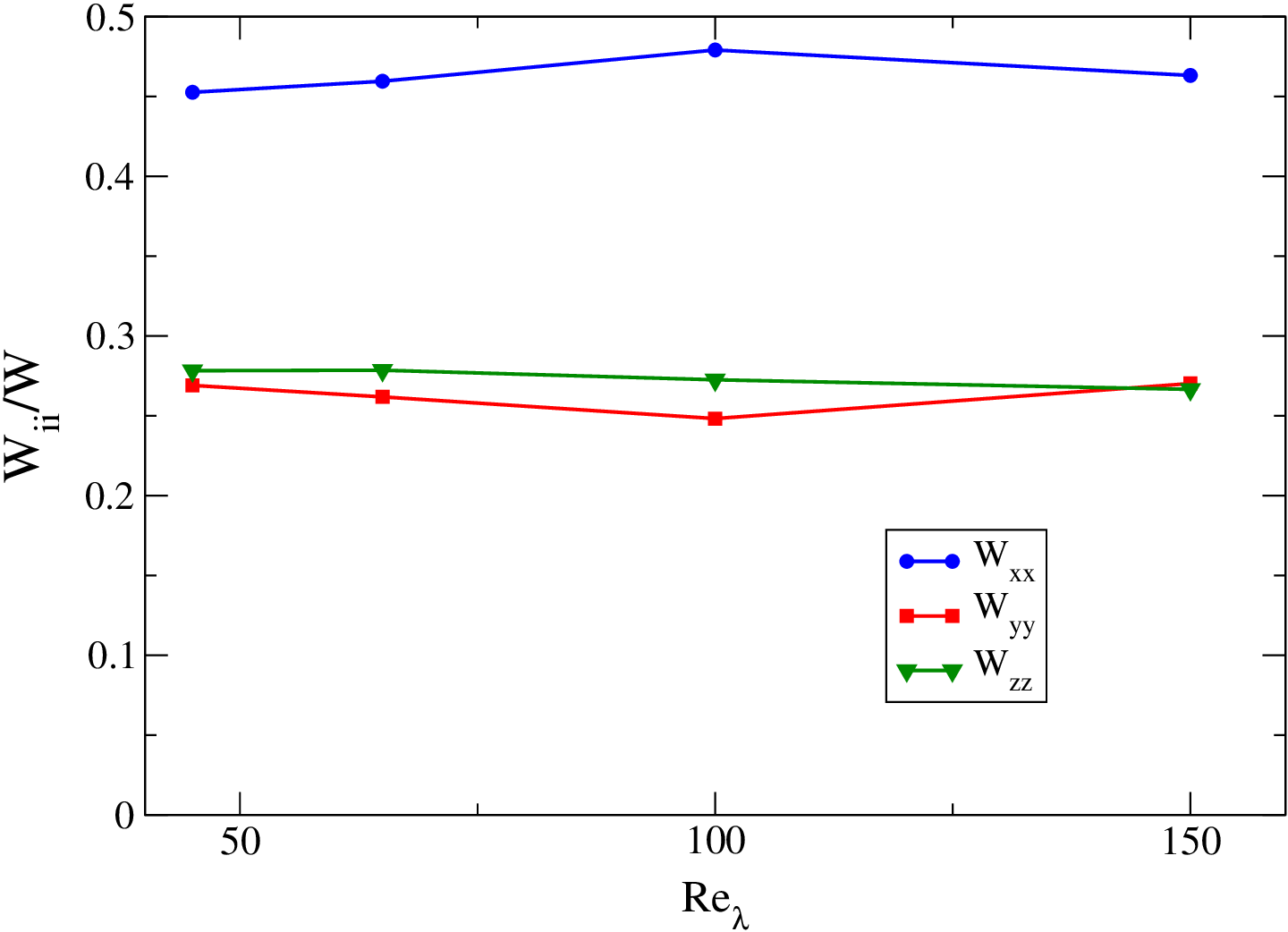}
\caption{\label{f:LES}   Relative components of the kinetic energies   
$K_{i}/K $  for  for constant-shear flow.
Results of the LES~\cite{LES}. Lines serve only to guide the eye.}
\end{figure}

\begin{figure}
 \centering\includegraphics[width=0.48 \textwidth]{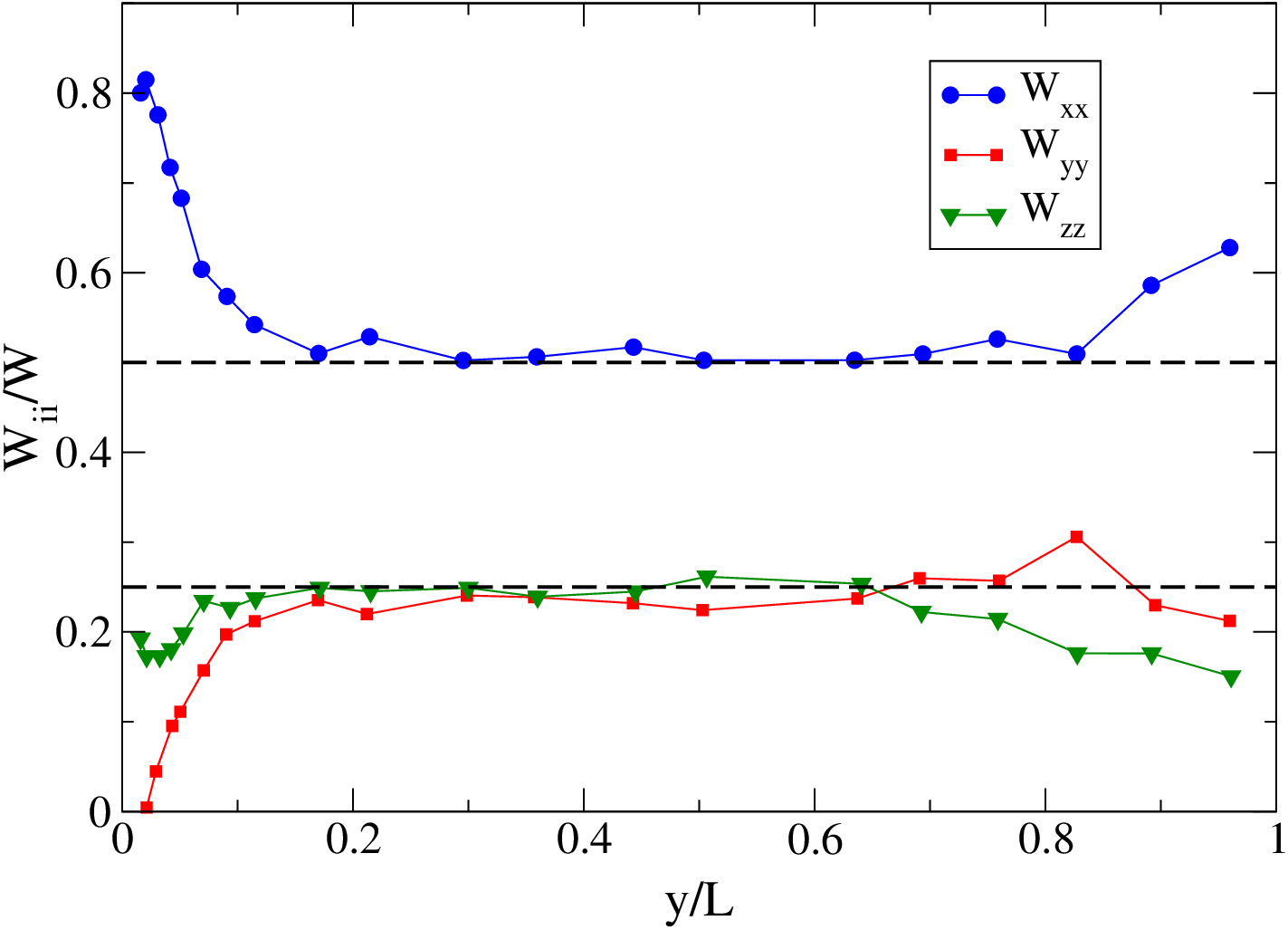}
\caption{\label{f:exp} Experimental profiles of the relative kinetic 
energies   in the vertical water channel with  $\RE =1000$ according to 
Ref.~\cite{Exp}. Solid lines serve  only to guide the eye.  Dashed lines 
show the  energy distribution~(\ref{Ksplit}).}
\end{figure}

The anisotropy of turbulent boundary layers, characterized by the 
dimensionless ratios $K_j/K$,
plays an important role in various phenomena and was a subject of 
experimental and theoretical interest for many decades, see, 
e.g.~\cite{79MY,Pope}.  Nevertheless, up to now the dispersion of 
results on this subject is too large. There is a widely spread opinion, 
based on old experiments, that the wall-normal turbulent fluctuations 
$K_y$ are much smaller than the other ones. For example, in the classical 
textbook by A.S. Monin and A.M. Yaglom~\cite{79MY} it was reported that 
in a neutrally stratified log-boundary layer $K_x=0.54\, K$, 
$K_y=0.06\, K$ and $K_z=0.40\, K$.  This appears to be in contradiction with simulation
results  at the largest Re$_\lambda$, $\RE$=590,  which are available in Ref.~\cite{DNS}. 
These results are reproduced in Fig.~\ref{f:DNS}.   Note the region of $\yp$ values,
about $100 < \yp < \frac23 \RE $, where the plots of $K_j/K  $  are 
nearly horizontal, as expected in the log-law region. From these plots 
we can conclude that is this region $K_x\approx 0.53\, K $ which is 
close to the ratio  $0.54$, stated in~\cite{79MY}. In contrast, the simulational
data for $K_y/K$ and $K_z/K$ are completely different. From 
Fig.~\ref{f:DNS} one gets $K_j\approx 0.22\,K$ and $K_z\approx 0.27\, K 
$.  Roughly speaking  $K_y$ is almost equal to  
$K_z$.   We propose that the difference between $K_y$ and $K_z$, observed 
in Fig.~\ref{f:DNS}, is due to the effect of the spatial energy flux, that 
is expected to vanish  in the asymptotic limit $\RE\to \infty$, 
but is still present at  $\RE$ values available in direct numerical simulations~\cite{DNS}. Indeed, for 
both  $\RE$ shown in Fig.~\ref{f:DNS} $K_y=K_z$ in the center of the 
channel, where the energy flux vanishes by symmetry. Clearly, there is 
no spatial energy flux also in a homogeneous constant shear turbulent flow.  Based on the similarity 
between the wall bounded and constant shear turbulent flows we expect 
the  values  of $\Kp_i$  to be the same for both flows in the limit 
$\RE\to \infty$.  The expectation  $K_y=K_z$ is confirmed by  
Large Eddy Simulation (LES) of  a constant shear flow~\cite{LES}. As 
one sees in Fig.~\ref{f:LES} in this flow $K_x \approx 0.46\, K$, while  
$K_y \approx K_z\approx 0.27\, K$.
For both flows  $K_z\approx 0.27\, K$, while in the 
stream-wise and and the wall-normal directions there is some difference in 
the ratios $K_x/K$ (0.53 vs. 0.46) and  $K_y/K$  (0.22 vs. 0.27).
We believe that these differences are again finite \Re~effects. This viewpoint is 
supported by a  recent laboratory
experiment by A. Agrawal, L. Djenidi and R.A. Antonia~\cite{Exp} in
a vertical water channel with  Re$_\lambda=1000$ which is reproduced in  
Fig.~\ref{f:exp}. Indeed  the experimental  values of $K_x/K$, 
$K_y/K$ and  $K_z/K$  distribute similarly to  
channel simulations and constant-shear LES:
\BE\label{Ksplit}
K_x=  K/2 \,, \qquad K_y=K_z= K/4 \ .
\EE

The operational conclusion of this discussion is that there are experimental
and simulational grounds to believe that for both constant-shear and wall-bounded 
turbulent flows in the log-law region the turbulent kinetic 
energy is distributed in a very simple manner: the stream-wise 
component contains a half of total energy,  the rest is equally 
distributed between wall-normal and cross-stream components. We propose that
Eq. (\ref{Ksplit}) is asymptotically exact.

\noindent\textbf{III. Structure functions  in isotropic,
 constant-shear, and wall bounded flows.} In isotropic 
homogeneous turbulence  the second order velocity structure functions, 
$S_2(\r)$, [Eq.~(\ref{LS2})] and $\~S_2(\r)$ [Eq.~(\ref{TS2})] are 
very well studied. They are invariant to the direction of the separation 
vector $\B r$ and up to intermittency corrections they read
\begin{equation}
S_2(r) =C_2(\epsilon r)^{\zeta_2}\,, \quad  \~S_2(r) =\~C_2(\epsilon 
r)^{\zeta_2}\,, \quad \zeta_2= 2/3\ .  \label{S2iso}
\end{equation}
The ratio of the 
dimensionless constants $\~C_2/C_2=4/3$ follows from the incompressibly 
constraint, while the values of   $\~C_2$, $C_2$ are known from 
extensive experiments and simulations\cite{Fri,79MY}:
\BE\label{C2}
\~C_2= 4\,C_2/3\,, \qquad C_2\approx 2.0\ .
\EE

For $r$ larger than the outer scale of turbulence, the correlations between 
velocities in $r$-separated points vanishes and the structure 
functions~(\ref{S2iso})  saturate at their asymptotic values:
\BE\label{sat1}
S_2(r)\to 2 \big< u_{\parallel}^2\big>\,, \quad \~S_2(r)\to 2 \big< 
u_{\perp}^2\big>\  .
\EE 
It is useful to define crossover scales $\ell_2$ and $\~\ell_2$  for the 
longitudinal and transversal structure functions as  follows:
\BE\label{sat2}
S_2(\ell_2)= 2 \big< u_{\parallel}^2\big>\,,
\quad \~S_2(\~\ell_2)= 2 \big< u_{\perp}^2\big>\ .
\EE
Clearly, in isotropic turbulence $\big< u_{\parallel}^2\big>=\big< 
u_{\perp}^2\big>$ and therefore the scales  $\ell_2$ and $\~\ell_2$ are    
related: 
\BE\label{rel1}
\~ \ell_2=\big( C_2 \big /  \~C_2 \big)^{3/2}\ell_2=  3\sqrt{3} \, 
\ell_2/ 8\approx 0.65 \,\ell_2\ .
\EE

The issue of the structure functions in anisotropic turbulent flows is
considerably  more involved \cite{99ALP,05BP}. In addition to the isotropic contribution~(\ref{S2iso}), the structure 
functions have anisotropic components, $S_{2,j}$, belonging to
irreducible representations of the SO(3) group with $j\ne 0$,
\begin{equation}
S_2(\r) = \sum_{j=0}^\infty \sum_{m=-j}^jS^{(2)}_{j m} (\r)
\end{equation}

The various anisotropic sectors exhibit $\ell$ dependent  scaling exponents:
\BE \label{z2ell}
S^{(2)}_{j m} (\r)\propto r^{\zeta^{(2)}_j}Y_{ j m}(\r/r)\,,
\EE
where the angular behavior of $S^{(2)}_{j,m} (\r)$ is carried by the spherical 
harmonic  $Y_{ j m}(\r/r)$.  The leading correction to the isotropic
sector ~(\ref{S2iso}) is given the $j=2$ contribution 
with $\zeta^{(2)}_2 \approx \frac 43$.  The mixture of contributions with 
different different exponents, each with an amplitude  that depends on the 
distance to the wall, may give the false impression that the scaling exponents of 
$S_2$ and $\~S_2$ are different; this impression disappears once the structure
functions are projected on the various sectors of the symmetry group, where their
exponents are the same (appearing universal) , but their amplitudes are of course non-universal, 
and see full details in \cite{05BP}.

Importantly,   all the anisotropic contributions~(\ref{z2ell}) 
can be  eliminated  by averaging  over  all the directions of $\r$ because
\BE\label{ever}
\int Y_{j m}\( \r / r \) d  (\r / r) =0\,, \quad \mbox{for}\
j \ne 0\ .
\EE
After the elimination of the anisostropic sectors, the  isotropic parts of $S_{2}(\r)$, $\~ S_{2}(\r)$,
i.e.  $S_{2,0}(\r)$, $\~ S_{2,0}(\r)$ exhibit the same scaling behavior as 
structure functions in isotropic turbulence, even in relatively low $\RE$ channel flows~\cite{99ABMP}.
The situation for high $\RE$ constant-shear and wall-bounded turbulence 
is even simpler. As demonstrated by Eq.~(\ref{Ksplit}), in these flows 
only the streamwise direction $\x$ is  special, the partial 
kinetic energies in two other directions are equal: $K_z=K_y$. Therefore 
in the limit $\RE\to \infty$ the symmetry of these two flows  can be 
considered as axisymmetric with the axis in the $\x$ direction, and one can
eliminate the anisotropic contributions by averaging only in the 
$(\x,\z)$-plane, parallel to the wall.

This idea is supported by experiments in the atmospheric turbulent boundary 
layer~\cite{00KLPS}),  in which the ground normal velocity component  $u_y$ was
measured at point of fixed height $y$ above the ground, separated by $\B \rho$ which was parallel to the
ground. Denote the resulting structure function as $\~S_{2}(y,\B \rho)$.
After averaging over the azimuthal angle $\phi$ in the   $(\x,\z)$-plane, this function
exhibits  homogeneous scaling behavior~(\ref{S2iso}) for all scales up to separations 
$\rho$ in the $(\x,\z)$-plane which are close to the wall distance $y$:
\BE\label{S2scaling}
\< \~S_{2}(y,\rho)\>_\phi  =  \~ C_2 (\epsilon \rho 
)^{2/3}\,, \quad 
\rho\le y\ .
\EE
The constant $\~C_2$ here is the same as 
in isotropic turbulence. Inspired by this  experiment (and see Fig.~3 in 
Ref.~\cite{00KLPS}) we make the assumption that  \emph{the 
largest of the two crossover scales (\ref{sat2}), namely  $\ell_2$, is 
determined by the distance to the wall}:
\BE\label{as1}
 \ell_2=y\ .
 \EE
 In other words, the assumption is that the structure function $\< \~S_{2}(y,\rho)\>_\phi$ changes sharply from a scaling law in $\rho$ to its asymptotic constant value precisely at $\rho=y$. This assumption introduces an unknown factor of the order of unity to our
 arguments; we take this factor to be exactly unity. \\
 
\noindent\textbf{Relationship between  $C_2$,  $c\N$ and  $\kappa\K$}.
We have presented all the ingredients necessary to estimate  the von K\'arm\'an constant. This constant will be related to  the Kolmogorov constant $C_2$ which appears in homogeneous isotropic flows and to the ratio $c\N$ which appears in homogeneous constant shear flows.
Using Eqs,~(\ref{Ksplit}), (\ref{sat2}), (\ref{rel1})  
and  (\ref{as1}) in Eq.~(\ref{S2scaling}) one gets:
\BE\label{der1}  K= 4K_y=2 
\<u_y^2\>=\~C_2(\epsilon\~\ell_2)^{2/3}=C_2(\epsilon y)^{2/3}\ .
\EE
In wall units $ \Kp= C_2(\epsilon^+\yp)^{2/3}$.
  Taking $\epsilon$ from Eq.~(\ref{input}) and $\Kp$ from 
Eq.~(\ref{res1}) we have  
\BE\label{der3}
c\N^{-2}=C_2\, (\Sp \Wp \yp)^{2/3}\ .
\EE
Together with Eqs.~(\ref{shear+}) and (\ref{bal3}) this leads to the 
relationship:
 \BE\label{final}
\kappa\K= \(\, c\N \sqrt {C_2}\, \)^3\ .
\EE
 Using the experimental values $c\N\approx 0.53$ and $C_2\approx 2.0$ we get 
$\kappa\N\approx 0.42$ 
in excellent agreement with the known value of this constant, 
$\kappa\N\approx 0.44\pm 0.03$. It should be stressed that in fact we have used 
only one assumption~(\ref{as1}) about the cross-over scale of the
structure function. All the other input is taken from  homogeneous 
data without any wall in sight. We propose that the
numerical agreement with known value of $\kappa\K$  reflects the quality of the  input values of 
$C_2$ and $c\N$.   

\acknowledgements
We thank Carlo M. Casciola for sharing with us the results of his LES. This work had been
supported in part by the US-Israel Bi-national science foundation and the European Commission
under a TMR grant.

\end{document}